\documentclass[useAMS,usenatbib,onecolumn]{mn2e}
\usepackage{subfigure}
\usepackage{rotating}
\usepackage{times}
\usepackage{graphicx}
\usepackage{lscape}
\usepackage{psfig}
\def\etal{{\it et al.}}
\def\swift{{\it Swift }}
\def\sqiggt{\hbox{\rlap{\lower.55ex \hbox {$\sim$}}\kern-.05em \raise.4ex \hbox{$>$}\,}}
\def\sqiglt{\hbox{\rlap{\lower.55ex \hbox {$\sim$}}\kern-.05em \raise.4ex \hbox{$<$}\,}}

\title[Can X-ray emission powered by a spinning-down magnetar explain some GRB light curve features?]{Can 
X-ray emission powered by a spinning-down magnetar explain some GRB light curve features?}
\author[N. Lyons \etal ]{N. Lyons$^{1}$\thanks{E-mail:
nal14@star.le.ac.uk}, P.T. O'Brien$^{1}$,
B. Zhang$^{2}$, R. Willingale$^{1}$, E. Troja $^{1,}$ $^{3,}$ $^{4}$, R.L.C. Starling$^{1}$\\
$^{1}$Department of Physics \& Astronomy, University of Leicester,
University Road, Leicester, LE1 7RH, UK\\
$^{2}$Department of Physics \& Astronomy, University of Nevada Las Vegas,
4505 Maryland Parkway, Box 454002, Las Vegas, NV 89154-4002, USA \\
$^{3}$INAF - Istituto di Astrofisica Spaziale e Fisica Cosmica, Sezione di Palermo,
 via Ugo la Malfa 153, 90146 Palermo, Italy \\
$^{4}$Dipartimento di Scienze Fisiche ed Astronomiche, Sezione di Astronomia, 
Universit`a di Palermo, Piazza del Parlamento 1,90134 Palermo, Italy}

\begin{document}

\date{Accepted 00. Received 00; in original form 00}

\pagerange{\pageref{firstpage}--\pageref{lastpage}} \pubyear{000}

\maketitle

\label{firstpage}

\begin{abstract}

Long duration gamma-ray bursts (GRBs) are thought to be produced by
the core-collapse of a rapidly-rotating massive star. This event
generates a highly relativistic jet and prompt gamma-ray and X-ray
emission arises from internal shocks in the jet or magnetised
outflows. If the stellar core does not immediately collapse to a black
hole, it may form an unstable, highly magnetised millisecond pulsar,
or magnetar. As it spins down, the magnetar would inject energy into
the jet causing a distinctive bump in the GRB light curve where
the emission becomes fairly constant followed by a steep decay when
the magnetar collapses.  We assume that the collapse of a massive star
to a magnetar can launch the initial jet. By automatically fitting the
X-ray lightcurves of all GRBs observed by the
\swift\ satellite we identified a subset of bursts which have a feature 
in their light curves which we call an internal plateau --- unusually
constant emission followed by a steep decay --- which may be powered
by a magnetar. We use the duration and luminosity of this internal
plateau to place limits on the magnetar spin period and magnetic field
strength and find that they are consistent with the most extreme
predicted values for magnetars.

\end{abstract}


\section{Introduction}

GRBs are thought to be caused by a violent event such as the collapse
of a massive star (for long duration bursts) or the coalescence of two
compact objects (for short duration bursts). These progenitors result
in the immediate formation of a black hole which powers a relativistic 
jet pointing in the direction of the observer. In the standard fireball model
variability in the Lorentz factor of the outflow causes internal shocks
which produce the prompt flash of X-ray and gamma-ray emission 
(Rees \& M\'esz\'aros 1994; Sari \& Piran 1997). When
the relativistic outflow sweeps up a sufficient amount of external
material, the ejecta is decelerated causing a forward shock which is
primarily responsible for the multi-wavelength afterglow emission
(Katz 1994; M\'esz\'aros \& Rees 1997;Sari, Piran \& Narayan 1998).

Alternatively there is a model that suggests a black hole may not be
formed immediately, but instead that a transitory highly magnetised
rapidly rotating pulsar, or magnetar, may form (Usov 1992; Thompson
1994), before the star collapses to a black hole (Rosswog \&
Ramirez-Ruiz 2003). Proto-magnetars have very high magnetic field
strengths of $10^{16}$G (Duncan \& Thompson 1992; Duncan 1998) which
are thought to be a consequence of millisecond rotation at birth in a
core-collapse supernova. Values up to $\sim 10^{17}$G are implied by
observations (Stella \etal\ 2005).  Such objects are considered a
possible central engine for GRBs due to their large rotational energy
reservoir, E$_{rot}$. Also they can be associated with supernovae, as
are long GRBs, and their winds are thought to become relativistic like
a GRB jet.

Zhang \& M\'esz\'aros (2001) investigated the observational signature
of a spinning-down magnetar as the GRB central engine. Adopting an
approximate magnetic dipole radiation model, they infer that the
spindown power of the proto-magnetar could produce a period of
prolonged constant luminosity followed by a $t^{-2}$ decay. They
considered the modification of the forward shock dynamics by magnetar
spindown and predicted a distinct achromatic feature. A similar model
was discussed earlier by Dai \& Lu (1998) who considered the energy
injection to the forward shock by a millisecond pulsar with much a
weaker magnetic field. This model is one of the candidates to
interpret the majority of the X-ray plateaus observed in many Swift GRB
afterglows (Zhang et al. 2006; Nousek et al. 2006). This model does
not invoke the internal dissipation of the magnetar wind. On the other
hand, if the magnetar wind indeed dissipates internally before hitting
the blastwave, it is possible that it would generate an "internal"
X-ray plateau whose X-ray luminosity tracks the spindown luminosity if
the energy dissipation and radiation efficiency remain constant. If
the magnetar undergoes direct collapse into a black hole before spin
down, then the X-ray plateau would be followed by a very steep
decay. This is the light-curve feature we investigate and 
hereafter we call this feature an "internal plateau".

\begin{figure}
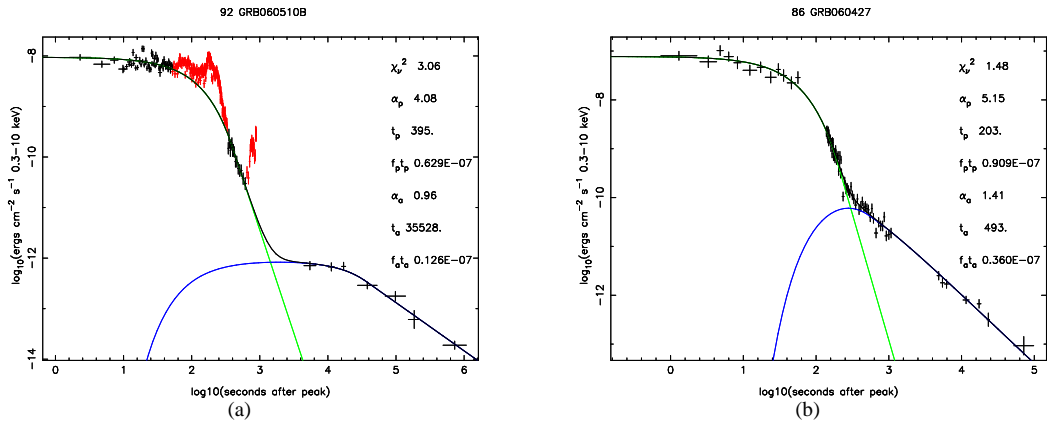

\centering
\subfigure[]
  {
   \includegraphics[width=5.2cm,angle=270]{plot_GRB060510B.ps}
     }
  \hspace{1.0cm}
\subfigure[]
 {
   \includegraphics[width=5.2cm,angle=270]{plot_GRB060427.ps}
 }
\caption{The left panel shows the light curve in the BAT and XRT for 
GRB 060510B and the right panel displays a more typical burst; GRB 060427. 
The green line represents emission from the burst (prompt) and the
blue line emission from the afterglow, as given by the Willingale et
al. (2007) model. The portions in red in the left panel are the 
data (flares and internal plateaus) which the model does not fit.}
\label{fitexamples}
\end{figure}

In the \swift\ era the early X-ray light curve, observed within the
first few hours of the GRB, has been found to be complex
(e.g. Nousek \etal\ 2006; O'Brien \etal\ 2006). The so-called
canonical X-ray light curve observed in a significant fraction of 
GRBs (Evans \etal\ 2009) has a short period of fast decay, 
often with flares superimposed, which are usually over within the first
hour after the burst.  This is followed by a shallower decay period
lasting from a few hours up to a day with a temporal decay index
$\alpha \sim 0.5$ (where the X-ray flux, $f_{\nu} \propto \nu^{-\beta}
t^{-\alpha}$ and $\beta$ is the spectral index). After this X-ray
plateau there is a smooth transition to a modest power law decay of 
$\alpha \sim 1-1.5$.

Willingale \etal\ (2007) found that the X-ray light curve of most
GRBs, including those of the canonical form, can be represented by two
components -- the prompt and afterglow --- plus flares (Section
2). However, we find that in a small minority of bursts a period of
relatively constant emission (compared to the general lightcurve)
followed by a steep decline can be identified which does not fit this
phenomenological model. The observed feature instead resembles the
proposed signature of a magnetar spin down. Troja \etal \ (2007) found
such a feature dominates the X-ray light curve of GRB 070110 from
$\approx 1,000 - 20,000$ seconds in and proposed it was due to a
spinning down millisecond pulsar. Starling \etal\ (2008) found a
similar, earlier feature in GRB 070616 ending at about 600
seconds. Liang et al. (2007) systematically analyzed a sample of X-ray
plateaus and identified several more plateaus that are followed by
decays with slopes steeper than $\alpha = -3$. It is the combination
of a plateau followed by a steep decay which distinguishes these from
the canonical behaviour. We regard these objects as candidate internal
plateaus.

We have conducted a systematic investigation of the GRB X-ray light
curves observed by \swift\ up to the end of 2008. Using an automated
fitting procedure in this paper we identify 10 bursts which may have an
emission component powered by magnetar spin-down dominating the light
curve for some period of time in the form of an internal plateau.
Assuming this internal plateau is caused by the spinning down of a
magnetar we use its properties to constrain the magnetic field and
initial spin period of the magnetar. The criteria for selecting those
GRBs with internal plateaus is discussed in Section 2.  Section 3
compares the properties of the internal plateaus with the magnetar
model and we discuss the implications in Section 4.

\section[]{The functional form of Swift Light curves}

Over 90\% of GRB X-ray light curves are well described by a two component
model with a prompt and an afterglow component as described in 
Willingale \etal\ (2007). These components are described by an exponential 
that relaxes into a power law; this function can be expressed for the prompt
component as shown in Equation 1. 
Large flares were masked out of the fitting procedure. Although
apparently bright, such flares account for only about 10\% of the
total fluence in most cases. The prompt emission rises with the time
constant $t_{p}$ and later the emission transitions from an
exponential to a power law at point (T$_p$, F$_p$), where the
subscript $p$ refers to the prompt component of the emission
(Willingale \etal\ 2007). The exponential and power law decay are both
controlled by the index $\alpha_{p}$. 

\begin{equation}
\begin{array}{l l l}
f_{p}(t)= & F_{p} {\rm exp}\left( {\alpha}_{p}-
\frac{t\alpha_{p}}{T_{p}} \right) {\rm exp}\left( \frac{-t_{p}}{t}
\right), & \quad t<T_{p} \\
\\
f_{p}(t)= & F_{p} \left( \frac{t}{T{_p}}\right)^{-{\alpha}_{p}} {\rm
exp}\left( \frac{-t_p}{t}\right) , & \quad t \geq T_{p} \\
\end{array}
 \end{equation}

For this investigation we are interested in those GRBs whose early
X-ray emission could not be adequately fitted by the Willingale
model. Thus we fitted all the \swift\ GRBs with the model and then examined
all cases where the model fails.

The X-ray light curves were derived from the BAT and XRT data using 
the methods described in O'Brien \etal\ (2006) and Willingale \etal\ (2007).  
BAT \& XRT light curves were derived for each GRB using the NASA's HEASARC
software. The BAT data were extracted over the 15-150 keV band and the
BAT spectra were produced using the task batbinevt. An estimate of the
fractional systematic error in each BAT spectral channel from the BAT
calibration database (CALDB) was added to the spectra using the
batphasyserr command. The corresponding response matrices were
generated by the command batdrmgen.
XRT data were extracted over the 0.3-10 keV band. The light curves
were corrected for Point Spread Function (PSF) losses and exposure
variations. The XRT spectra were extracted using the xselect software
and the spectra was grouped to have at least 20 counts per
channel. The relevant ancillary response files were generated using
the task xrtmkarf .
UVOT data were extracted using the task uvotevtlc. The V, B, U and
white magnitudes were corrected for Galactic extinction along the line
of sight and then converted to monochromatic fluxes at the central
wavelength of each filter.  The effective mid-wavelength was taken to
be 4450\AA\ for the White filter.

In Fig.~\ref{fitexamples} we show an example of a GRB that the Willingale
model fits well (GRB 060427) and one which it does not (GRB 060510B)
and instead demonstrates an internal plateau where the flux remains
constant with small fluctuations for about 360 seconds. It has been
suggested that instead of a plateau there is a group of flares very
close together, however maintaining a high level of flux for hundreds
of seconds with the peak of each flare having almost identical flux
seems unlikely. Some of the GRBs for which the
model fails were those where a pre-cursor triggered the \swift\
BAT instrument or where very large flares were not fitted. We examined
all of the fitted light curves to remove such cases. 
To be included in the internal plateau sample a GRB must have

\begin{enumerate}
\item A lightcurve that could not be adequately fitted by the Willingale model.
\item A significant period of time during which the X-ray flux is relatively constant, i.e. at least a third of 
a decade long.
\item A convincing steep decline following the internal plateau which falls by a factor of ten where $\alpha$ $\geq$ 4, so
that the emission is likely caused by central engine activity and is not the canonical X-ray plateau.
\end{enumerate}

This gives 10 GRBs with light curve internal plateau features that
resemble the spin-down magnetar model discussed in Zhang \&
M\'esz\'aros (2001).

\section[]{GRBs with an internal plateau}

\begin{figure}
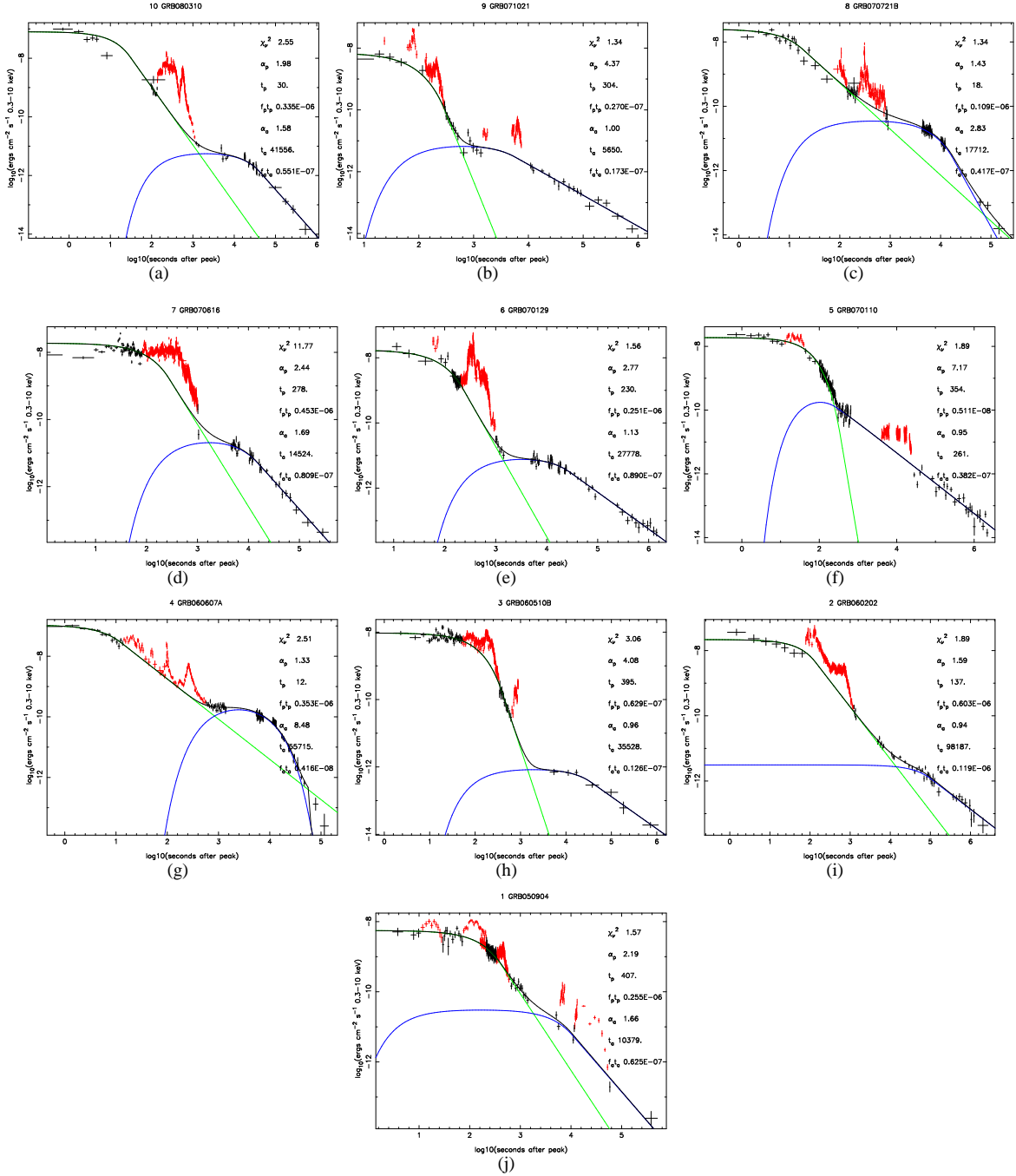

\flushleft
\subfigure[]
 {
  \includegraphics[width=4.0cm,angle=270]{GRB080310fit.ps}
}
\centering
\subfigure[]
  {
   \includegraphics[width=4.0cm,angle=270]{GRB071021fit.ps}  
  }
  \hspace{0.5cm}
\subfigure[]
 {
  \includegraphics[width=4.0cm,angle=270]{GRB070721Bfit.ps}  
}

\subfigure[]
{
 \includegraphics[width=4.0cm,angle=270]{GRB070616fit.ps} 
}
\subfigure[]
 {
  \includegraphics[width=4.0cm,angle=270]{GRB070129fit.ps}
 }  
\subfigure[]
 {
  \includegraphics[width=4.0cm,angle=270]{GRB070110fit.ps}
 } 
 \subfigure[]
 {
  \includegraphics[width=4.0cm,angle=270]{GRB060607Afit.ps}
} 
\subfigure[]
{
 \includegraphics[width=4.0cm,angle=270]{GRB060510Bfit.ps}
} 
\subfigure[]
{
 \includegraphics[width=4.0cm,angle=270]{GRB060202fit.ps}
} 
\subfigure[]
{
 \includegraphics[width=4.0cm,angle=270]{GRB050904fit.ps}
}  
\caption{The GRB light curves displaying internal plateau behaviour.
The green line represents emission from the burst (prompt) and the
blue line emission from the afterglow, as given by the Willingale et
al. (2007) model. The portions in red are the data (flares and
internal plateaus) which the model does not fit.}
\label{sample}
\end{figure}

The 10 GRBs which form our internal plateau sample and values of interest
such as the redshift and plateau limunosity are listed in Table 1. For
the GRBs with an observed redshift in Table 1, the mean redshift is 3.96,
significantly higher than the \swift\ mean redshift of 2.22 for all
GRBs with measured redshift. A K-S test with a confidence level of
90$\%$ could not prove the that the distribution of these redshifts
are inconstient with the \swift\ redshift distribution for all GRBs
with measured redshift. The GRBs which display an internal plateau are
shown in Fig.~\ref{sample} and are discussed briefly below.

GRB 080310 has emission that could be an internal plateau followed by
a steep decline, which seems to rise above the underlying
emission. Also shortly after the internal plateau there is a flare
which peaks at the same flux as the internal plateau.  While in this
GRB the proposed internal plateau could be due to a multiple number of
flares (O'Brien \etal\ in preparation), we include it in our sample.

GRB 071021 has a possible internal plateau dominating the early X-ray
lightcurve. This is the shortest proposed internal plateau in the
sample lasting about 105 seconds.

GRB 070721B has a possible internal plateau that dominates early in
the lightcurve. Flaring dominates over the internal plateau emission,
during the middle of this time interval. This could signify a brief
period of accretion onto the proto-magnetar. Ignoring the single flare
the emission is similar to that for other internal plateau candidates,
so it has been included in the sample. 

GRB070616 is intriguing in that the emission rises relatively slowly
over 100 seconds to a peak, then persists at a fairly constant level
before showing a very rapid decline.

GRB 070129 is similar to GRB 070721B in that it has a possible
internal plateau that is interrupted by a flare followed by a steep
decline.

GRB 070110 displays a canonical early light curve with an initial
steep decline, but then exhibits a period of relatively constant
emission. Following this plateau the decay is surprisingly steep
($\alpha \sim 7$) decay (Troja \etal\ 2007). Thus in this case the
proto-magnetar survived for much longer than in most of the other GRBs.

GRB 060607A appears to follow the canonical lightcurve with a "normal"
X-ray plateau with multiple flares preventing a good fit with the two
component model. However at late times the decay following the plateau
is too steep for an afterglow and is consistient with $\alpha$ $\sim
4$. This is unlikely to be explained by anything other than central
engine activity and thus has been included in the internal plateau
sample. As in GRB 070110, the internal plateau seen in GRB 060607A
dominates the burst emission unusually late starting at about 900 seconds
when (from Table 1) most of the other internal plateaus have ended.

GRB 060510B (also shown in Fig.~\ref{fitexamples}) is very similar to
GRB 070616. In both cases the proposed internal plateau dominates the
emission from the burst very early on.

GRB 060202 displays unusual emission attributed to an internal plateau
between 325 and 766 seconds. The fluctuations during this plateau are
unusually regular.

GRB 050904 has multiple flares at early and late times, but at about
230 seconds there is a period where the emission appears relatively
constant followed by a steep decay, leading it to be included
in the sample as a possible internal plateau.

\begin{table}
\caption{The observed properties of the GRBs with an internal plateau.}
\begin{tabular}{| c | c | c | c | c | c |}
\hline
GRB & Redshift & Flux $^{1}$ & Luminosity $^{1}$ & End Time ${^1}$ & Steep Decay \\
    &          & $10^{-9}$ erg cm$^{-2}$ s$^{-1}$ 0.3-10 keV & erg s$^{-1}$ & s  \\
\hline
 080310  & 2.426 & 5.39 & 2.6e+50 & 401.9 & $11.21^{+1.00}_{-0.50}$ \\
 071021  & 5.0 & 2.45 & 6.6e+50 & 248.3 & $9.18^{+1.01}_{-0.474}$ \\
 070721B & 3.626 & 0.24 & 3.0e+49 & 802.9 & $10.31^{+1.42}_{-2.696}$ \\
 070616  & 2.22$^{*}$ & 11.44 & 4.4e+50 & 585.6 & $5.07^{+0.13}_{-0.17}$ \\
 070129  & 2.22$^{*}$ & 2.24 & 8.6e+49 & 683.0 & $7.71^{+0.88}_{-0.67}$  \\
 070110  & 2.352 & 0.02 & 8.8e+47 & 21887.1 & $6.98^{+0.10}_{-0.34}$  \\
 060607A  & 3.082 & 0.15 & 1.3e+49 & 13294.7 & $3.43^{+0.80}_{-0.91}$ \\
 060510B & 4.9 & 6.58 & 1.7e+51 & 362.9 & $10.43^{+0.66}_{-0.58}$ \\
 060202  & 2.22$^{*}$ & 2.69 & 1.0e+50 & 766.0 & $5.70^{+0.17}_{-0.16}$ \\
 050904  & 6.29 & 1.53 & 7.1e+50 & 488.8 & $9.364^{+0.91}_{-1.49}$ \\
\hline
\end{tabular}
\\
$^{*}$ Where no measurement is available
the redshift is assumed to be the mean of \swift\ GRBs 
 taken from the website maintained by
P. Jakobsson \\ $http://raunvis.hi.is/~pja/GRBsample.html$ \\
$^{1}$ These are parameters related to the plateau, i.e. the 
end time is the time the plateau ends before the steep decline begins.
\end{table}

\vspace{1cm}

\begin{figure}
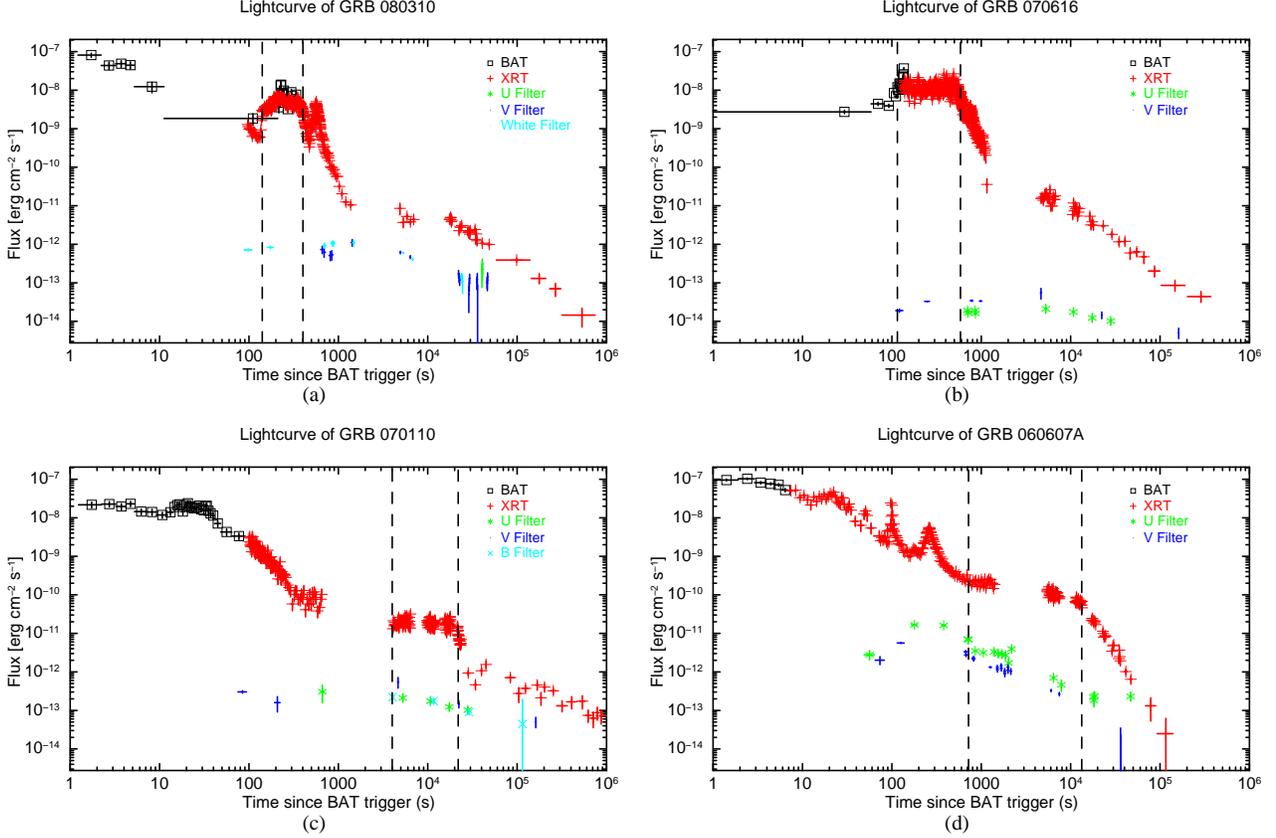

\centering
\subfigure[]
 {
  \label{test2}
  \includegraphics[width=5.2cm,angle=270]{UVOTlc080310.ps}
}
\label{optical1}
\centering
\subfigure[]
  {
   \label{test3}
   \includegraphics[width=5.2cm,angle=270]{UVOTlc070616.ps}  
  }
  \hspace{0.5cm}
\subfigure[]
 {
  \label{test4}
  \includegraphics[width=5.2cm,angle=270]{UVOTlc070110.ps}  
}
\subfigure[]
 {
  \label{test5}
  \includegraphics[width=5.2cm,angle=270]{UVOTlc060607A.ps} 
}
\caption{Combined BAT, XRT, and UVOT light curves
for the 4 GRBs with multi-wavelength data during the internal plateau. 
The vertical dashed lines indicate the time interval 
over which the internal plateu dominates the emission. The optical
data have been scaled down by a factor of 10 for Figure 3a and c.}
\label{optical2}
\end{figure}

To further investigate the nature of the internal plateau we compared
the X-ray data to optical/UV data from the UVOT.
The GRBs within the sample with near-simultaneous optical/UV and X-ray
light curves are shown in Fig. ~\ref{optical2}. While an early rise in
the optical can be seen in some cases, the optical emission does not
show the same behaviour as the X-ray. The internal plateau and
following steep decay are significantly more prominent in X-rays.  For
example in GRB 070616, the optical is constant from before the plateau
in the X-ray and until after the steep decline.

In Fig. ~\ref{sample} if the plateau seen in each of the X-ray
lightcurves is of an external origin, then the X-ray and optical
lightcurve should be related to each other in a manner consistent with
the external shock model, i.e. the breaks should be
achromatic. However, if the X-ray and optical emission components are
not related to each other, e.g. a sharp decay in X-ray but no break in
optical, this strongly suggests that the X-ray emission is not
external or a jet-break but rather is of internal origin.


In Troja \etal\ (2007) for GRB 070110 four spectral energy
distributions (SEDs) were examined during the initial decay, the
beginning and end of the plateau and during the shallow decay after
the steep decline. These SEDs were constructed by extrapolating the
X-ray spectrum to the lower energies. During the initial decay the
optical data are not consistent with the extrapolation of the X-ray
spectrum to low energies. During the internal plateau, the optical and
X-ray spectral distributions are also completely inconsistent with one
another, implying different origins for the optical and X-ray
photons. 

For GRB 080310 and GRB 070616 the extrapolation of the X-ray spectrum
is also inconsistient with the optical during the internal plateau
(Beardmore \etal\ in preparation, Starling \etal\ 2008).  Likewise,
for GRB060607A extrapolating the X-ray spectum to the optical in a
similar way to Troja \etal\ (2007) gives a poor fit to the optical
(reduced $\chi^{2}$ of 15.1). From this we conclude that during the
internal plateau, the X-ray and optical emission have separate origins
for the four GRBs for which we have multi-wavelength data. Henceforth
we concentrate on the X-ray behaviour of our sample.

As the time at which the internal plateau ends differs markedly ({\it
cf.} GRB 070110 and GRB 070616) it is possible they have a different
origin. Thus we further sub-divide the sample into those GRBs in which the
constant emission phase ends before or after 10,000 seconds. Those
which end before 10,000 seconds are denoted as having early internal
plateaus whereas those ending after this time have late internal
plateaus.
In the next section we compare the results for these two groups to
determine if their properties are consistent with being caused by the
same physical process and whether that process is consistent with
being due to a magnetar. Of the 8 GRBs in the early internal
plateau group 5 have a redshift measurement as do both GRBs in the late
plateau group. For the 3 GRBs with no redshift measurement we adopt a redshift
of 2.22, the mean redshift of \swift GRBs to determine the
luminosity. Our conclusions are not sensitive to this choice.

\section[]{The magnetar model}

In order to generate the intense magnetic fields required for a
proto-magnetar a massive star's magnetic field must be increased as it
collapses through magnetic-flux conservation or efficient dynamo
action (Dai \& Lu 1998). This can be used to make a prediction for the
initial period of the proto-magnetar; every time the star collapses
inwards by a factor of two the magnetic fields are increased by a
factor of four. To build up sufficient dynamo action on the surface
the star needs an initial rotation period of $\leq$ 10ms (Usov
1992). Another method to predict the shortest rotation period is to
use the breakup spin-period for a neutron star, which is $\geq$ 0.96ms
Lattimer \& Prakash (2004). The inital rotation period of milliseconds
are thought to differentiate between a proto-magnetar and a neutron
star. From a theoretical estimate the limits set for the expected
strong magnetic field are B $\geq 10^{15}$G (Thompson 2007).

To place limits on the central object we assume the GRB jet is
launched by the collapse of a massive star to a magnetar which
survives for a short period of time before it collapses to a black
hole (see Thompson 2007 for a review on the magnetar GRB central
engine models). A transitory proto-magnetar could cause the flux to
remain roughly constant throughout the plateau until the
proto-magnetar had spun-down enough for the rotational energy to be
insufficient to support the star. It would then collapse to form a
black hole ceasing the plateau-like emission and causing the steep
decay following the plateau. Flares during the plateau-like emission
or the steep decline can arise from accretion onto the central object.
We use equations 2 and 3 (see Zhang \& M\'esz\'aros., 2001) to relate
the continuous injection luminosity of the plateau, $L$, and the
rest-frame time at which the plateau breaks down, $\tau$, to the
magnetar magnetiec field and initial period.

\begin{equation}
 L \simeq 10^{49} B_{p,15}^{2} P_{0,-3}^{-4} R_{6}^{6} \qquad $erg$ \ $s$^{-1}
\end{equation}

\begin{equation}
\tau = 2.05 \times 10^{3} I_{45} B_{p, 15}^{-2} P_{0,-3}^{2} R_{6}^{6} \qquad $s$
\end{equation}

\begin{figure}
\center
\psfig{file=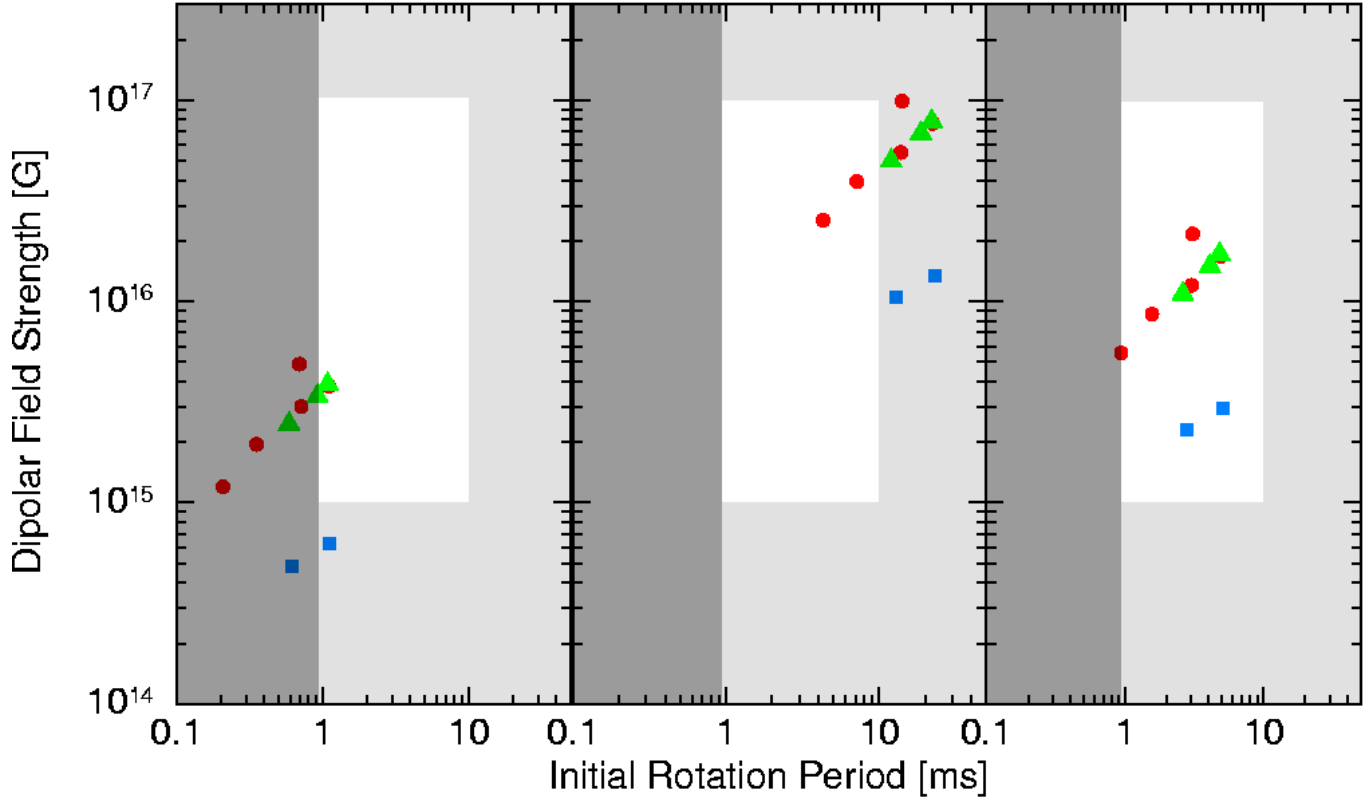,width=18cm}
\caption{The initial period and magnetic field for each of the 
GRBs examined. In the left-hand panel it was assumed that energy was
released isotropically, whereas in the middle and right-hand panels it
is beamed with an opening angle of 4 and 18 degrees respectively. GRBs
with red filled circles have known redshifts and their internal
plateaus occur during the prompt emission; GRBs shown by blue filled
squares have known redshifts and their internal plateaus occur after
the prompt emission; GRBs shown by green filled triangles have
internal plateaus that occur during the prompt emission at unknown
redshifts, and for which the redshift has been assumed to be equal to
the median redshift of the \swift sample, meaning their parameters are
more uncertain. The light grey shaded regions show limits based on the  
magnetic field and period limits discussed in the literature. See 
text for details. The darker grey shaded region shows where a progenitor 
would be violating the breakup spin-period of a neutron star. }
\label{periodfield}
\end{figure}

We use the GRB spectral shape and a k-correction (Bloom \etal\ 2001)
to convert the observed 0.3--10 keV flux to the rest-frame
1--1,000~keV luminosity.  $B_{p}$ is the magnetic field strength at
the poles where $B_{p,15} = B_p / 10^{15}$ G, $P_{0,-3}$ is the
initial rotation period in milliseconds, $I_{45}$ is the moment of
inertia in units of $10^{45}$ g cm$^{2}$ and $R_{6}$ is the stellar
radius in units of $10^{6}$ cm. If we use standard values for a
neutron star (Stairs 2004) of mass $\sim 1.4 M_{\odot}$ and $R_{6}
\sim 1$ then using Equations 2 and 3 we can infer the central object's
initial rotation period and magnetic field strength.  The correlation
between the derived period and the magnetic field is shown in
Fig.~\ref{periodfield}.

\begin{figure}
\center
\includegraphics[width=5.5cm,angle=270]{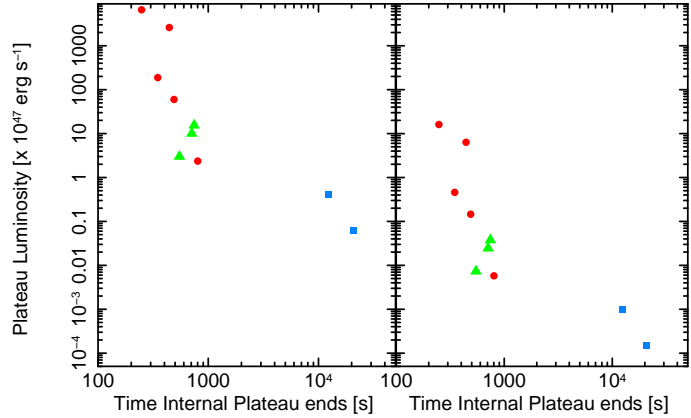}
\caption{The relationship between the length of the internal plateau
emission and its luminosity in the observers frame, 
where it was assumed that energy was released isotropically in the
left panel and beamed with an opening angle of 4 degrees in the right
panel. GRBs with red filled circles have known redshifts and their
internal plateaus occur during the prompt emission, GRBs shown by blue
filled squares have known redshifts and their internal plateaus occur
after the prompt emission. The GRBs shown by green filled triangles
have internal plateaus that occur during the prompt emission at
unknown redshifts.}
\label{timeluminosity}
\end{figure}

In theory there should be GRBs in the lower right portion of
Fig.~\ref{periodfield} with a relatively long period and low dipolar
field strength. From Equation 2, a lower luminosity is expected for
these GRBs and hence it may be the internal plateau is too faint to be
observable. GRBs are unlikely to be present in the top-left as they
would require extreme magnetic fields.

The derived periods are close to the sub-millisecond break-up limits
for a neutron star, so it could be that most stars cannot support a
temporary magnetar and collapse immediately to a black hole. If the
initial rotation of the proto-magnetar was violated by the break-up
limit for a neutron star's period it is unlikely it could become
stable enough to survive for the lengths of time given in Table
1. This results in a natural boundary on the left-side of
Fig.~\ref{periodfield}. Thus only a small group of GRBs may produce an
observable plateau and this could explain the apparent correlation in
Fig.~\ref{periodfield}.

The rotational energy reservoir of the magnetar given in Table 2 was
calculated using Equation 4 with R$_{6} = 1$ and is consistent with
the total power of the internal plateau ($E_{iso, plat}$) as it should
be given the way magnetic field and initial period are calculated.

\begin{equation}
E_{rot} = 2 \times 10^{52} M_{1.4} R_{6}^{2} P_{0,-3}^{-2} \qquad $ergs$
\end{equation}

The plateau energy, $E_{\gamma, iso}$, was calculated assuming that
the radiation is emitted isotropically but it is almost certainly
collimated by a relativistic wind flowing through a cavity produced by
the elongation of a bubble of plasma and magnetic field (Bucciantini
\etal\ 2007). This can be corrected for using

\begin{equation}
E_{\gamma} = f_{b} \times E_{\gamma, iso}	\qquad $where$ f_{b}=(1-cos\theta_{j}) = 0.5 \times \theta_{j}^{2}
\end{equation}

where $\theta_{j}$ is the opening angle of the beam. The maximum
beaming angle ($\theta = 18^{\circ}$) was estimated by assuming the
fastest possible period as the break up spin-period of a neutron
star. Taking this angle as the beaming angle for each GRB, the
corresponding beaming-corrected energies are shown in Table 2 along
with an example of the beaming-corrected energies derived using a
beaming angle of 4 degrees (Frail \etal\ 2001). A factor which effects
the comparison of these energies is that the true initial rotation
period is likely to be smaller than that derived from Equation 4
(Thompson 2007), so $E_{rot}$ could be larger.

\begin{table}
\caption{The different beamed energies found for the plateau for different opening angles compared to the energy of the actual GRB and the energy availiable in the rotational
energy resevoir. All energies in Table 2 are in ergs, the opening
angles used to find the beamed energy respectively are 4 and 18
degrees. The E$_{iso}$ values were derived from lightcurves with the
0.3-10 keV band}
\begin{tabular}{| l | l | l | l | l | l | l | l | l | l |}
\hline
GRB & Isotropic $P_{0}$ & Isotropic $B_{p}$ & Beamed $P_{0}$ & Beamed $B_{p}$ & $E_{iso} $ & $E_{rot}$ & $E_{iso, plat}$ & $E_{\gamma1, plat}$ & $E_{\gamma2, plat}$ \\
    & ms & $\times 10^{16}$G & ms & $\times 10^{16}$G &            &          &                  & $\theta_{j} = 4$ & $\theta_{j} = 18$ \\
\hline
 080310  & 0.7 & 0.3 & 13.8 & 5.5 & 2.91e+53$^{1}$ & 3.90e+52 & 4.00e+52 & 9.75e+49 & 5.29e+51 \\
 071021  & 0.7 & 0.5 & 14.1 & 9.9 & 3.53e+53$^{2}$ & 4.15e+52 & 4.25e+52 & 1.04e+50 & 8.45e+51 \\
 070721B & 1.1 & 0.4 & 22.3 & 7.7 & 1.72e+53$^{3}$ & 1.64e+52 & 1.69e+52 & 4.11e+49 & 1.27e+51 \\
 070616  & 0.6 & 0.2 & 12.0 & 5.0 & 2.47e+54$^{4}$ & 5.74e+52 & 2.93e+53 & 7.13e+50 & 1.50e+52 \\
 070129  & 0.9 & 0.3 & 18.7 & 6.9 & 3.98e+53$^{5}$ & 2.34e+52 & 1.07e+53 & 2.62e+50 & 5.51e+51 \\
 070110  & 1.1 & 0.06 & 23.2 & 1.3 & 7.08e+52$^{6}$ & 1.61e+52 & 1.65e+52 & 4.02e+49 & 1.07e+51 \\
 060607A & 0.6 & 0.04 & 12.4 & 1.0 & 2.13e+53$^{7}$ & 5.35e+52 & 5.48e+52 & 1.34e+50 & 8.89e+51 \\
 060510B & 0.2 & 0.1 & 4.2 & 2.5 & 1.09e+54$^{8}$ & 4.64e+53 & 4.76e+53 & 1.16e+51 & 1.78e+52 \\
 060202  & 1.1 & 0.4 & 22.0 & 6.9 & 3.08e+53$^{9}$ & 1.70e+53 & 8.33e+52 & 2.03e+50 & 4.27e+51 \\
 050904  & 0.4 & 0.2 & 7.1 & 4.0 & 2.51e+54$^{10}$ & 1.61e+53 & 1.65e+53 & 4.01e+50 & 3.17e+52 \\
\hline
\end{tabular}
\\
$^{1}$ Tueller \etal\ 2008
$^{2}$ Barbier \etal\ 2007
$^{3}$ Palmer \etal\ 2007
$^{4}$ Sato \etal\ 2007
$^{5}$ Krimm \etal\ 2007
$^{6}$ Cummings \etal\ 2007
$^{7}$ Tueller \etal\ 2006
$^{8}$ Barthelmy \etal\ 2006
$^{9}$ Hullinger \etal\ 2006
$^{10}$ Sakamoto \etal\ 2005
\end{table}

The correlation between plateau luminosity and duration is shown in
Fig.~\ref{timeluminosity}, which suggests that higher luminosity
plateaus are generally of shorter duration. There are too few GRBs in
the late internal plateau group to draw any firm conclusions. Their
luminosities are lower, but not much lower than that of the early
internal plateau group.

\section[]{Discussion}

We have identified a small number of GRBs which display a period of
time during which the X-ray emission shows a smooth plateau followed
by a steep decline. The internal plateau is challenging to interpret
using accretion models as it requires a constant power jet component
with a roughly constant radiation efficiency. This possibility has
been examined by Kumar \etal\ (2008a), who suggest that the prompt
emission of a GRB may be caused by the accretion of the outer regions
of a stellar core and that the X-ray plateau could be caused by the
fall-back and accretion of the stellar envelope. This model has
problems accounting for the steep declines seen after the
plateau. Even assuming a sharp edge to the region being accreted, the
steepest decline expected is $\alpha \sim 2.5$ (Kumar et al. 2008b).

Here we argue that a more natural explanation may come from the
magnetar model which predicts a period of constant spin-down
power. This model starts with the assumption that the neutron star
accretor can power the GRB prompt emission which while not certain, is
feasible (Usov 1992; Thompson 1994; Bucciantini \etal\
2007). Comparison of the luminosity and duration of the internal
plateaus observed in our GRB sample with the dipolar spindown law
(Zhang \& M\'esz\'aros 2001) implies upper limits to the magnetic
field strengths close to the maximum allowed for such objects and
initial spin periods also close to the maximum allowed to maintain
neutron star structural integrity. The upper limits for the dipolar
magnetic field of the magnetar are particularly strong if the emission
is strongly beamed.

The largest magnetic fields implied for isotropic emission are
consistent with field strengths of $\times 10^{16}$G which can be
generated in magnetars born with spin of a few milliseconds (Thompson
\& Duncan 1993; Duncan 1998). A giant flare from SGR 1806-20 on
$27^{th}$ December 2004 demonstrated that unless such flares are much
rarer than the rate implied by detecting one, magnetars must possess a
magnetic field strength of $\sim 10^{16}$G or higher. Indeed values up
to $\sim 10^{17}$G could not be ruled out (Stella \etal\ 2005). For
the GRB sample in this paper this could allow beaming factors
corresponding to jet opening angles of 4-10 degrees, consistent with
values derived from Frail
\etal\ (2001).

The number of GRBs that display internal plateau behaviour is 
very small. This perhaps is not surprising as we would expect them to
only be detectable for quite a narrow combination of magnetic field
strength and initial spin period.  These rare features do provide
limits on the magnetic fields surrounding the central engine around
the GRB, and can help advance understanding of the mechanisms behind
prompt emission.

\section[]{Acknowledgements}

We gratefully acknowledge funding for \swift\ at the University of Leicester by the Science and Technology Facilities Council, 
in the USA by NASA and in Italy by contract ASI/INAF I/088/06/0. NL and RLCS also acknowledge funding by STFC via a studentship and PDRA respectively. 
BZ and NL acknowledge funding by NASA grants NNG05GB67G and NNX08AN24G. We are also very grateful to our colleagues on the \swift\ project 
for their help and support, particularly Kim Page for providing the BAT \& XRT lightcurves.

\section[]{References}

Barbier L. et al. 2007, GCN 6058 \\
Barbier L. et al. 2007, GCN 6966 \\
Barthelmy S. et al. 2006, GCN 5107\\
Bloom, J., S., Frail, D., A., Sari, R. 2001, ApJ, 121, 2879 \\
Bucciantini, N. et al. 2007, MNRAS, 380, 1541 \\
Cummings, J., R. et al. 2007, GCN 6007 \\
Dai, Z., G., Lu, T. 1998, A \& A, 333, L87 \\
Duncan R., C., 1998, ApJ, 498, L45 \\
Evans et al. 2009, MNRAS, submitted \\
Frail, D. A. et al. 2001, ApJ, 562, 55 \\
Hullinger D. et al. 2006, GCN 4635 \\
Katz, J. I. 1994, ApJ, 422, 248 \\
Krimm H. et al. 2007, GCN 6058 \\
Kumar, P., Narayan, R., Johnson, J., L. 2008a, A \& A, 388, 1729 \\
Kumar, P., Narayan, R., Johnson, J., L. 2008b, Science, 321, 376 \\
Lattimer, J., M., Prakash, M. 2004, Sci, 304, 536 \\
Liang, E., Zhang, B., Zhang, B. 2007, ApJ, 670, 565 \\
M\'esz\'aros, P., Rees, M. J. 1997, ApJ, 476, 232 \\
Mundell, C. et al. 2007, ApJ, 642, 389 \\
Nousek, J., A. et al. 2006, ApJ, 642, 389 \\
O'Brien, P.T. et al. 2006, ApJ, 647, 1213 \\
Palmer D. et al. 2007, GCN 6643 \\
Panaitescu, A. et al. 2006, MNRAS, 369, 2059 \\
Rosswog, S., Ramirez-Ruiz E., 2003, AIPC, 727, 361 \\
Rees, M. J., \& M\'esz\'aros, P. 1994, ApJ, 430, L93 \\
Sakamoto, T. et al. 2005, GCN 3938 \\
Sari, R., \& Piran, T. 1997, ApJ, 485, 270 \\
Sari, R., Piran, T., Narayan, R. 1998, ApJ, 497, L17 \\
Sato G. et al. 2007, GCN 6551 \\
Stairs, I. H. 2004, Science, 304, 547 \\
Starling, R.L.C., et al. 2008, MNRAS, 384, 504 \\
Stella, L. et al. 2005, ApJ, 634, L165 \\
Thompson, C. \& Duncan, R., C. 1993, ApJ, 408, 194 \\
Thompson, C. 1994, MNRAS, 270, 480 \\
Thompson, T., A. 2007, RMxAC, 27, 80 \\
Troja, E. et al. 2007, 665, 599 \\
Tueller J. et al. 2006, GCN 5242 \\
Tueller J. et al. 2008, GCN 7402 \\
Usov, V., V. 1992, Nature, 357, 472 \\
Willingale R. et al., 2007, ApJ, 662, 1093 \\
Zhang B., M\'esz\'aros P., 2001, ApJ, 552, L35 \\
Zhang, B. et al. 2006, ApJ, 642, 354 \\


\label{lastpage}

\end{document}